\documentclass[aps,preprint,showpacs,nofootinbib,12pt]{revtex4}
\usepackage{graphicx}% Include figure files
\usepackage{epsfig}
\usepackage[dvips]{color}

\begin{document}

\title{ The decay rate of $\psi(2S)$ to $\Lambda_c+\overline{\Sigma^+}$ in SM and beyond}

\author{ Hong-Wei Ke$^{1}$   \footnote{khw020056@hotmail.com},
         Ya-Zheng Chen$^{2}$ \footnote{cyzufo@mail.nankai.edu.cn} and
         Xue-Qian Li $^{2}$  \footnote{lixq@nankai.edu.cn} }

\affiliation{
  $^{1}$ School of Science, Tianjin University, Tianjin 300072, China \\
  $^{2}$ School of Physics, Nankai University, Tianjin 300071, China
  }

\begin{abstract}
\noindent With rapid growth of the database of the BES III and the
proposed super flavor factory, measurement on the rare $\psi(2S)$
decays may be feasible, especially the weak decays into baryon
final states. In this work we study the decay rate of $\psi(2S)$
to $\Lambda_c+\overline{\Sigma^+}$ in the SM and physics beyond
the SM (here we use the unparticle model as an example). The QPC
model is employed to describe the creation of a pair of $q\bar q$
from vacuum. We find that the rate of $\psi(2S)\rightarrow
\Lambda_c+\overline{\Sigma^+}$ is at order of $10^{-10}$ in the
SM,  whereas the contribution of the unparticle is too small to be
substantial. Therefore if a large branching ratio is observed, it
must be due to new physics beyond SM, but by no means the
unparticle.

\end{abstract}

\pacs{13.20.Gd, 12.39.Jh}

\maketitle

\section{introduction}

It is well known that $\psi(2S)$ decays overwhelmingly via strong
and electromagnetic interactions, which are the OZI \cite{OZI}
suppressed processes. The goal of the research on $\psi(2S)$ is
not only to study the low energy QCD effects which determine the
hadronic transition matrix elements, but also to find evidence of
new physics in the concerned processes. Since the strong
interaction QCD dominates, the contributions of new physics whose
effects are in general much smaller, would be drowned by the QCD
background and unobservable. To observe the processes where new
physics may play more important role, we need to turn to the rare
decays of $\psi(2S)$. One obvious field is the weak decays of
$\psi(2S)$. Of course, due to the dominance of the strong
interaction at $\psi(2S)$ decays, the branching ratios of such
weak decays are very small, so that measurements on them are
extremely difficult.

In our earlier papers \cite{Wang} the branching ratios of
semi-leptonic and hadronic weak decays of J/$\psi$ into two mesons
(vector or pseudoscalar) are evaluated in the framework of the
standard model (SM) and we found that they are of order of
$10^{-9}$ to $10^{-11}$ which is much beyond the range of
detection for any present machines. However, the situation is not
so pessimistic, as the database accumulated at the BES III is
becoming large and the proposed super-flavor factory will be
running in the future, measurements on the weak decays may be
possible.  It is expected that if some sizable signals for the
weak decays are observed (even still very small), one would
conclude that there is new physics or new mechanism beyond our
present knowledge. Moreover, as stated above, the main goal is to
see if for some modes the ratios obviously exceed the values
theoretically calculated in terms of SM, if yes, it would be clear
signals for new physics beyond SM. That is what we expect.

For a long time more researches   focus  on mesons than baryons. The
reason is that there are three constituents  in baryon and there are
only two in mesons in comparison. The three-body system indeed is much
more complicated than the two-body one, even in the classical
physics. Another reason is that the data concerning baryon final
states are not as abundant as for mesons, so that it is harder to fix
the model parameters. But now a
larger database on baryon final states will be available at the
LHCb and other experiments such as BES-III, especially the proposed super-flavor factory. Since baryons possess
different characteristic from mesons a thorough study on the
production and decay processes of baryon is becoming an important field which
will provide a clearer insight into the structure of hadron,
underlying physics, mechanics and the fundamental interactions,
especially such processes may be more sensitive
to new physics beyond the SM. On the other hand we will also have a
chance to test some models and constrain their parameter space. Moreover, the plan of studying baryon case at BES III
is substantialized
% \footnote{We thank Dr. H.B. Li for the information and he suggested us to carry out such calculation
%which may be useful for the experimental search for the baryon final states via weak interaction at the BES III.}
, so that we would like to pay more attention to the weak decays
of $\psi(2S)$ into baryon final state, even though corresponding
processes are rare.

In this paper we study the production of  $ \psi(2S)\rightarrow
\Lambda_c+\overline{\Sigma^+}$ which occurs via weak interaction.
Concretely, we will calculate the rate in the frameworks of the SM
and beyond. As an example for the model beyond the SM, we employ
the unparticle scenario proposed by Georgi \cite{Georgi:2007ek}.
Some phenomenology concerning the unparticle scenario has been
explored \cite{Cheung:2007zza, Unparticle} and constraints on its
parameter space were set by fitting observational data. Therefore,
the concerned model parameters have been determined and there is
not much room for adjusting them. In this work we will estimate
the contributions from both the SM particles and the unparticle as
the intermediate agent to $\psi(2S)\rightarrow
\Lambda_c+\overline{\Sigma^+}$. A comparison between their
contributions may shed some light on the role of the possible
models beyond the SM.

Since three pairs of quarks and antiquarks are produced from
either the intermediate bosons or vacuum, the processes are more
complex than the case for mesons. But it does not cause any
principal difficulty. In this work we will employ the
Quark-Pair-Creation model(QPC
model)\cite{Micu:1968mk,yaouanc-book,Ping:2004sh} to deal with the quark pair production and the harmonic
oscillator wave function for the hadrons. The quark-antiquark pair
is created from vacuum in this model and the essence of the
picture is that the quark-pair is created from the soft gluons
radiated from the quark-antiquark legs.

The key point is to evaluate the transition matrix elements between
hadrons which are governed by the non-perturbative QCD. The
harmonic oscillator wave function is used to describe the bound
states in this model and the corresponding non-perturbative effect is
included in the model-parameter $R$ .

This paper is organized as follows: after the introduction, in
section II and  III we formulate the transition amplitude of the
process $\psi(2S)\rightarrow \Lambda_c+\overline{\Sigma^+}$ in SM
and unparticle model respectively. The last section is devoted to
our conclusion and discussions.

\section{$\psi(2S)\rightarrow
\Lambda_c+\overline{\Sigma^+}$ in SM}

In this section we are going to calculate the decay widths of
$\psi(2S)\rightarrow \Lambda_c+\overline{\Sigma^+}$ in SM.

%\begin{figure}
%\begin{center}
%\begin{tabular}{cc}
%\includegraphics[width=9cm]{LFQM.eps}
%\end{tabular}
%\end{center}
%\caption{ Feynman diagram for meson transition amplitude}
%\label{fig:LFQM}
%\end{figure}

\subsection{Feynman diagram}\label{formula}
\begin{figure}
\begin{center}
\begin{tabular}{ccc}
\includegraphics[scale=0.5]{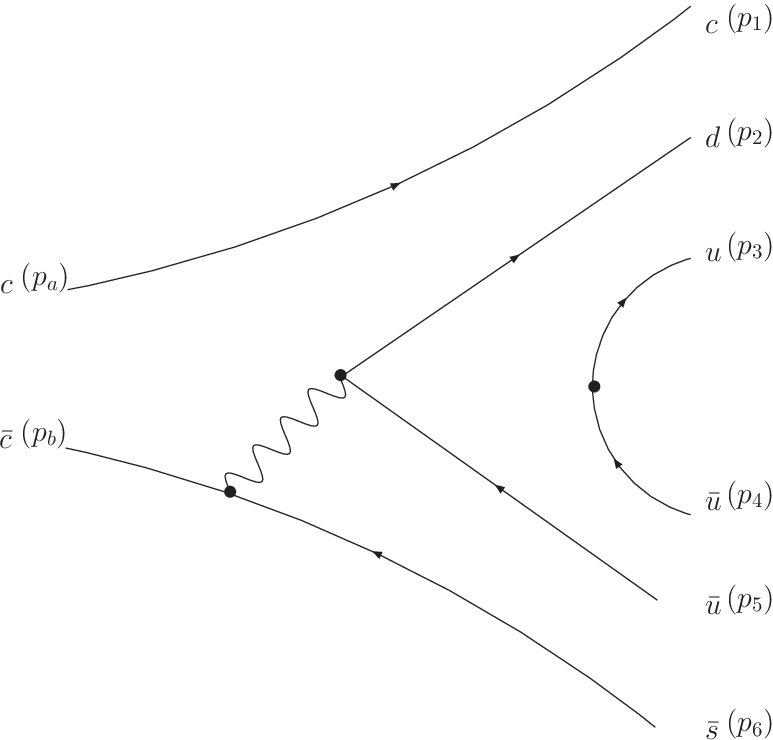}
\qquad\includegraphics[scale=0.5]{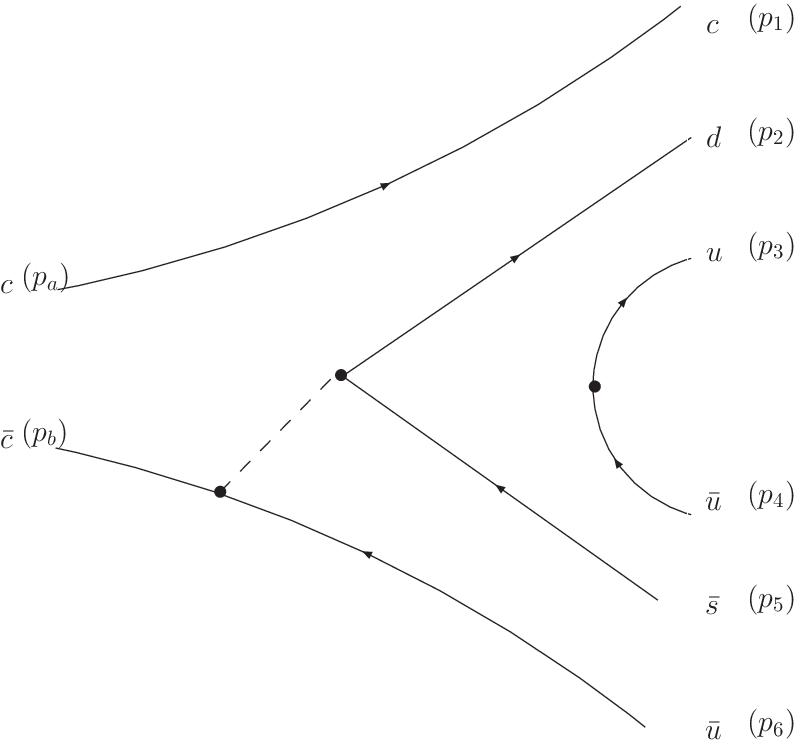}
\\(a)\qquad\qquad\qquad\qquad\qquad\qquad\qquad\qquad\qquad\qquad(b)\\
\end{tabular}
\end{center}

\caption{The Feynman diagrams of $\psi(2S)\rightarrow
\Lambda_c+\overline{\Sigma^+}$. (a) in SM; (b) beyond SM (the
dotted line is the unparticle propagator) } \label{Fig:QPC}
\end{figure}

We show the Feynman diagrams in Fig.\ref{Fig:QPC} and the left one
is that in the SM. The c quark directly transits into the finial
$\Lambda_c$ as a spectator  and $\bar c$ decays into $\bar s$ and
$d\bar u$ via a W-boson exchange. A pair of $u\bar u$ is created from the vacuum.
Then those quarks and anti-quarks eventually hadronize into
$\Lambda_c$ and $\overline{\Sigma^+}$.

\subsection{the QPC model}
In the Fig.\ref{Fig:QPC}  a pair of $u\bar{u}$ is created from
vacuum. The QPC model is also named as 3P0 model since the quantum
number of the vacuum is $J^{PC}=0^{++}$. In the QPC model the pair of
quark-antiquark is created from the vacuum while other quarks are
unaffected.

The S-matrix element is
\begin{eqnarray}\label{qpc1}
S_{fi}=\delta_{fi}-2\pi\delta(E_f-E_i)\langle f|H_I|i\rangle,
 \end{eqnarray}
and  the Hamiltonian $H_I$ which is responsible for the creation of
quark-antiquark pair from the vacuum\cite{yaouanc-book} is
\begin{eqnarray}
H_{I}&=&\int d \mathbf{p}_{q}d\mathbf{p}_{\bar q}[3\;\gamma
\delta^{3}(\mathbf{p}_{q}+\mathbf{p}_{\bar q})\sum_{m}\langle
1,1;m, -m|0,0\rangle
\nonumber\\&&\times\mathcal{Y}_{1}^{m}(\mathbf{p}_{q}-\mathbf{p}_{\bar
q})(\chi_{1}^{-m}\varphi_{0}\omega_{{ij}})]
b^{+}_{i}(\mathbf{p}_{q},s)d^{+}_{j}(\mathbf{p}_{\bar q},s'),
\end{eqnarray}
where  $\varphi_{0}=(u\bar u +d\bar d +s \bar s)/\sqrt 3 $ is an
SU(3) flavor-singlet,  $\omega_{ij}$ is a color-singlet,
$\chi_{1}^{-m}$ for a spin-triplet, $\mathcal{Y}_{1}^{m}$ is a solid
harmonic polynomial for the $L=1$ orbital angular momentum and
$\gamma$ is a dimensionless constant representing the strength of
the creation and generally it is a fully non-perturbative QCD factor
and must be determined by fitting data. The S-matrix element can be
expressed in terms of the amplitude $M$
\begin{eqnarray}\label{qpc2}
S_{fi}=\delta_{fi}-2\pi\delta^4(p_f-p_i)M.
 \end{eqnarray}

For a two-body  decay the width is
\begin{eqnarray}\label{qpc3}
\Gamma=2\pi\int
d\mathbf{p}_Bd\mathbf{p}_C\delta(m_A-E_B-E_C)\delta(\mathbf{p}_B+\mathbf{p}_C)|M|^2,
 \end{eqnarray}
where $A$ stands as the initial particle, B and C are the daughter
hadrons. It is noted that the Bjorken-Drell convention is used.

\subsection{Transition matrix element}
Now we begin to evaluate the transition matrix element in the QPC
model. First in terms of Eq.(\ref{qpc1}) we have
\begin{eqnarray}\label{qpc4}
T_{fi}=-2\pi\delta(E_f-E_i)\langle \Lambda_c
\overline{\Sigma^+}|H_WH_I|\psi(2S)\rangle,
 \end{eqnarray}
where the hamiltonian $H_W$ is responsible for the weak transition.
The expression becomes
\begin{eqnarray}\label{qpc5}
-i\frac{G_F}{\sqrt{2}}V_{cs}V_{ud}g^{\mu\nu}\bar{\psi_c}\gamma_\mu(1-\gamma_5)\psi_s
\bar{\psi_d}\gamma_\nu(1-\gamma_5)\psi_u(2\pi)^4\delta^4(p_b-p_6-p_5-p_2),
 \end{eqnarray}
and
\begin{eqnarray}\label{qpc6}
H_W=-i\frac{G_F}{\sqrt{2}}V_{cs}V_{ud}g^{\mu\nu}\bar{\psi_c}\gamma_\mu(1-\gamma_5)\psi_s
\bar{\psi_d}\gamma_\nu(1-\gamma_5)\psi_u\frac{(2\pi)^4}{m_\psi}\delta^3(p_b-p_6-p_5-p_2).
 \end{eqnarray}

In terms of  the wave functions listed in the Appendix we obtain
\begin{eqnarray}\label{qpc7}
T_{fi}=&&-2\pi\delta(E_f-E_i)\frac{\delta_{ab}}{\sqrt{3}}\varphi_\psi\chi_\psi
\frac{\varepsilon_{123}}{\sqrt{6}}\varphi_{\Lambda_c}\chi_{\Lambda_c}
\frac{\varepsilon_{456}}{\sqrt{6}}\varphi_\Sigma\chi_\Sigma\int
d\mathbf{p}_ad\mathbf{p}_bd\mathbf{p}_1d\mathbf{p}_2d\mathbf{p}_3d\mathbf{p}_4d\mathbf{p}_5d\mathbf{p}_6\nonumber\\&&
\langle L,S;M-m,m|J,M\rangle
\Psi^{tot}_J(\mathbf{p}_a,\mathbf{p}_b)\Psi^{tot}_{\Lambda_c}(\mathbf{p}_1,\mathbf{p}_2,\mathbf{p}_3)
\Psi^{tot}_{\Sigma}(\mathbf{p}_4,\mathbf{p}_5,\mathbf{p}_6)\nonumber\\&&
\langle 0|b_1b_2b_3d_4d_5d_6 \,\, H_WH_I\,\, b_a^\dag
d_b^\dag|0\rangle
\nonumber\\=&&-2\pi\delta(E_\psi-E_{\Lambda_c}-E_{\Sigma})\mathcal{F}_cC_{fs}\int
d\mathbf{p}_1d\mathbf{p}_2d\mathbf{p}_5
\psi_\psi\psi_{\Lambda_c}\psi_\Sigma\frac{1}{(2\pi)^6}\sqrt{\frac{m_c}{E_c}}\sqrt{\frac{m_s}{E_s}}\sqrt{\frac{m_u}{E_u}}
\sqrt{\frac{m_d}{E_d}}\nonumber\\&& \langle L,S;M-m,m|J,M\rangle
(-i\frac{G_F}{\sqrt{2}}V_{cs}V_{ud}g^{\mu\nu})\frac{(2\pi)^4}{m_\psi}
\bar{V_c}\gamma_\mu(1-\gamma_5)V_s\bar{U_d}\gamma_\nu(1-\gamma_5)V_u\nonumber\\&&
[3\gamma\Sigma_m\langle 1,1;m',-m'|0,0\rangle
\mathcal{Y}_1^m(2\mathbf{p}_{\Lambda_c}-2\mathbf{p}_1-2\mathbf{p}_2)]\delta^3(p_J-p_{\Lambda_c}-p_{\Sigma})
,
 \end{eqnarray}
where $\mathcal{F}_c$ is a color-related factor and $C_{fs}$ is a
flavor-spin factor.

\begin{eqnarray}\label{qpc8}
M=&&-\mathcal{F}_cC_{fs}\int
d\mathbf{p}_1d\mathbf{p}_2d\mathbf{p}_5
\psi_J\psi_{\Lambda_c}\psi_\Sigma\frac{1}{(2\pi)^6}\sqrt{\frac{m_c}{E_c}}\sqrt{\frac{m_s}{E_s}}\sqrt{\frac{m_u}{E_u}}
\sqrt{\frac{m_d}{E_d}}\nonumber\\&& \langle L,S;M-m,m|J,M\rangle
(-i\frac{G_F}{\sqrt{2}}V_{cs}V_{ud}g^{\mu\nu})\frac{(2\pi)^4}{m_\psi}
\bar{V_c}\gamma_\mu(1-\gamma_5)V_s\bar{U_d}\gamma_\nu(1-\gamma_5)V_u\nonumber\\&&
[3\gamma\Sigma_m\langle 1,1;m',-m'|0,0\rangle
\mathcal{Y}_1^m(2\mathbf{p}_{\Lambda_c}-2\mathbf{p}_1-2\mathbf{p}_2)].
 \end{eqnarray}
We need to emphasize that the expansion of the field operator $\psi$
and $\bar \psi$ and the anti-commutation relations of the creation
and annihilation operators need to follow the Bjorken-Drell
convention.

\section{$\psi(2S)\rightarrow
\Lambda_c+\overline{\Sigma^+}$ beyond SM}

Here we take the unparticle model as
 an examplefor the new physics beyond the SM. It is worth emphasizing that other new physics models beyond the SM
may behave quite differently from the unparticle and  their contributions to the mode may deviate by orders from that of unparticle.
The corresponding Feynman diagrams are presented in
Fig.\ref{Fig:QPC}(b) where the intermediate agent is replaced by an
unparticle. Unlike the W-boson, the unparticle is neutral and allows existence of a flavor-changing neutral curresnt, so $\bar
c$ decays into $\bar u$. We below will exploit the cases with two
kinds of unparticles i.e. scalar and vector unparticles.

For the scalar unparticle
\begin{eqnarray}\label{qpc7}
M=&&-\mathcal{F}_cC_{fs}\int
d\mathbf{p}_1d\mathbf{p}_2d\mathbf{p}_5
\psi_J\psi_{\Lambda_c}\psi_\Sigma\frac{1}{(2\pi)^6}\sqrt{\frac{m_c}{E_c}}\sqrt{\frac{m_s}{E_s}}\sqrt{\frac{m_u}{E_u}}
\sqrt{\frac{m_d}{E_d}} \langle L,S;M-m,m|J,M\rangle\nonumber\\&&
\frac{C_S^{cu}C_S^{sd}}{(\Lambda_\mathcal{U}^{d_\mathcal{U}})^2}\frac{A_{d_\mathcal{U}}}{2\sin(d_\mathcal{U}\pi)}
\frac{ip^\mu
p^\nu}{(-p^2)^{2-d\mathcal{U}}}\frac{(2\pi)^4}{m_\psi}
\bar{V_c}\gamma_\mu(1-\gamma_5)V_u\bar{U_d}\gamma_\mu(1-\gamma_5)V_s\nonumber\\&&
[3\gamma\Sigma_m\langle 1,1;m',-m'|0,0\rangle
\mathcal{Y}_1^m(2\mathbf{p}_{\Lambda_c}-2\mathbf{p}_1-2\mathbf{p}_2)],
 \end{eqnarray}
with $p$ is the momentum of the unparticle  and
\begin{eqnarray}
A_{d_\mathcal{U}}=\frac{16\pi^{\frac{5}{2}}}{(2\pi)^{2d_\mathcal{U}}}\frac{\Gamma(d_\mathcal{U}+\frac{1}{2})}{\Gamma(d_\mathcal{U}-1)
\Gamma(2d_\mathcal{U})}
\end{eqnarray}
where $d_\mathcal{U}$ is the scale dimension\cite{Georgi:2007ek}.

For the vector unparticle
\begin{eqnarray}\label{qpc8}
M=&&-\mathcal{F}_cC_{fs}\int
d\mathbf{p}_1d\mathbf{p}_2d\mathbf{p}_5
\psi_J\psi_{\Lambda_c}\psi_\Sigma\frac{1}{(2\pi)^6}\sqrt{\frac{m_c}{E_c}}\sqrt{\frac{m_s}{E_s}}\sqrt{\frac{m_u}{E_u}}
\sqrt{\frac{m_d}{E_d}} \langle L,S;M-m,m|J,M\rangle\nonumber\\&&
\frac{C_V^{cu}C_V^{sd}}{(\Lambda_\mathcal{U}^{d_\mathcal{U}-1})^2}\frac{A_{d_\mathcal{U}}}{2\sin(d_\mathcal{U}\pi)}
\frac{i(p^\mu
p^\nu/p^2-g^{\mu\nu})}{(-p^2)^{2-d\mathcal{U}}}\frac{(2\pi)^4}{m_\psi}
\bar{V_c}\gamma_\mu(1-\gamma_5)V_u\bar{U_d}\gamma_\mu(1-\gamma_5)V_s\nonumber\\&&
[3\gamma\Sigma_m\langle 1,1;m',-m'|0,0\rangle
\mathcal{Y}_1^m(2\mathbf{p}_{\Lambda_c}-2\mathbf{p}_1-2\mathbf{p}_2)].
 \end{eqnarray}

\section{Numerical results}
Since it is impossible to integrate out the variables
$\mathbf{p}_1$, $\mathbf{p}_2$ and $\mathbf{p}_5$ in $M$
analytically we instead proceed the whole calculation in
Eq.(\ref{qpc4}) numerically.

The input parameters   $m_u=m_d=0.33$GeV, $m_s=0.50$GeV,
$m_c=1.8$GeV and $\gamma=3$ are set in \cite{yaouanc-book}. The
parameter $R$ in the wave function is corresponding to $1/\beta$
in \cite{CCH2} which can be fixed   by fitting its decay constant.
Instead, by means of the binding energy $\Delta E$, we have
$R_{\psi(2S)}^2\approx 3$GeV$^{-1}$ \cite{yaouanc-book} which is
close to the number obtained by fitting the decay constant.
$R^2\approx6$GeV$^{-1}$\cite{yaouanc-book} was determined for
light baryon, thus we set $R_{\Sigma}^2=6$GeV$^{-1}$ in our
numerical computations. For $\Lambda_c$ if the $ud$ quarks can be
regarded as a diquark $R_{\Lambda_c}^2$ should be equal to
$R_{D_s}^2 ( i.e. \,\,\frac{1}{\beta_{D_s}^2})$ and $\beta_{D_s}$
can be found in Ref. \cite{CCH2}, approximately, but sufficiently
accurate, we set $R_{\Lambda_c}^2= 4$ GeV$^{-2}$. In our
calculation we suppose that the quarks which participate in the
weak transition are approximately on shell. The parameters in the
SM sector are $G_F=1.166\times 10^{-5}$, $V_{ud}=0.974$,
$V_{cs}=0.957$\cite{PDG08}. With all the parameters we get
$BR(\psi(2S)\rightarrow\Lambda_c+\overline{\Sigma^+})=2.87\times
10^{-10}$. If we replace $\bar s$ by $\bar u$ in Fig.\ref{Fig:QPC}
we obtain $BR(\psi(2S)\rightarrow\Lambda_c+\bar p)$ as $2.68\times
10^{-11}$.

For the unparticle scenario, according to the present literature,
the relevant parameters are fixed as $\Lambda_\mathcal{U}=1$TeV,
$d_\mathcal{U}=1.5$ $C_S^{cu}=0.021$, $C_S^{ds}$,
$C_V^{cu}=0.0005$ according to Ref.\cite{Li:2007by}, and we also
set $C_S^{ds}=C_S^{cu}$ and $C_V^{ds}=C_V^{cu}$ in our
calculations. If only the contribution from unparticle is
considered, one has
$BR(\psi(2S)\rightarrow\Lambda_c+\overline{\Sigma^+})$ to be
$5.35\times 10^{-25}$ for scalar unparticle and it is $2.40\times
10^{-19}$  for vector unparticle. Because they are many orders of
magnitude smaller than the contribution of the SM , the
interference between the SM and unparticle contributions are
negligible.

Thus it is found that  the branching ratio of baryonic decays
$\psi(2S)\rightarrow\Lambda_c+\overline{\Sigma^+}$ has the same
order of magnitude with that of $\psi(2S)$ decaying into  a
charmed meson plus something else\cite{Wang:2008tf}.

\section{Conclusion and discussion}
In this work we study the weak baryonic decay of $\psi(2S)$:
$\psi(2S)\rightarrow\Lambda_c+\overline\Sigma^+$ in the SM and
beyond. In the calculation we employ a phenomenological model i.e.
the QPC model to deal with the quark-antiquark pair creation and
for calculating the hadronic transition matrix elements we use the
harmonic oscillator wavefunction to describe both the baryonic and
mesonic bound states. The model parameters are fixed by fitting
the available experimental data. The advantage is that we can
reduce the model-dependence, while the disadvantage is that the
errors in the data will be unavoidably brought into the
theoretical predictions. As one has a better knowledge on the
error in the data, on the other side however, the uncertainties
are well controlled within a reasonable range, so the numerical
results are trustworthy, at least at the order of magnitude, even
though not very accurate.

The branching  ratio of
$\psi(2S)\rightarrow\Lambda_c+\overline{\Sigma^+}$ in the SM would
reach the order of magnitude as about $10^{-10}$, which is of the
same order as that for mesonic decays of $\psi(2S)$. Such small
branching fraction definitely cannot be observed by the present
experimental facilities.

At first look, the unparticle scenario might provide a larger
contribution to the decay mode because there does not exist a
heavy weak scale $M_W$ at its propagator. However, the the data of
$D^0-\bar D^0$ mixing\cite{Li:2007by}, neutrino
experiment\cite{Li:2007kj} and lepton flavor violation
\cite{Wei:2008wi} rigorously constrains the effective couplings of
the unparticle and the SM particles, thus one cannot freely adjust anything at all. The resultant
contribution of the unparticle to the decay width is many order of
magnitude smaller than that of the SM. Therefore this decay mode
cannot be a probe for the unparticle scenario which has been
explored by many authors.

The large database of $\psi(2S)$ which will be collected at BES
III and the will-be super-flavor factory may reach the accuracy to
realize the measurements on weak transitions. Just because of its
significance such experiment is required to be carried out.

Indeed, if such weak processes were observed to be
larger by orders than that we estimated above, it signifies existence of new
physics beyond the SM or new mechanisms or new hadronic structure, but by our
calculation, definitely not from the unparticle contribution.

\section*{Acknowledgments}

We thank Dr. Hai-Bo Li who suggested this research. This work is
supported by the National Natural Science Foundation of China
(NNSFC) under the contract No. 11075079 and No. 11005079; the
Special Grant for the Ph.D. program of Ministry of Eduction of
P.R. China No. 20100032120065.

\appendix

\section{Notations}

Here we list the wave functions for any meson and baryon
\begin{eqnarray}\label{app1}
&&|\psi\rangle=\frac{\delta_{ab}}{\sqrt{3}}\varphi\chi\int d
\mathbf{p}_ad \mathbf{p}_b \langle L,S;M-m,m|J,M\rangle
\Psi^{tot}_\psi |c(p_a) \bar{c}(p_b)\rangle\nonumber\\&&
|\Lambda_c\rangle=\frac{\varepsilon_{123}}{\sqrt{6}}\varphi\chi\int
d \mathbf{p}_1d \mathbf{p}_2 \mathbf{p}_3 \Psi^{tot}_{\Lambda_c}
|c(p_1)d(p_2)u(p_3)\rangle\nonumber\\&&
|\Sigma\rangle=\frac{\varepsilon_{456}}{\sqrt{6}}\varphi\chi\int d
\mathbf{p}_4d \mathbf{p}_5 \mathbf{p}_6 \Psi^{tot}_{\Sigma}
|\bar{u}(p_4)\bar{u}(p_5)\bar{s}(p_6)\rangle,
 \end{eqnarray}
where
$\Psi^{tot}_\psi=\delta(\mathbf{p}_a+\mathbf{p}_b-\mathbf{p}_\psi)\phi(\mathbf{p}_a-\frac{\mathbf{p}_\psi}{2},\mathbf{p}_b-\frac{\mathbf{p}_\psi}{2})$,
$\Psi^{tot}_{\Lambda_c}=\delta(\mathbf{p}_1+\mathbf{p}_2+\mathbf{p}_3-\mathbf{p}_{\Lambda_c})
\phi(\mathbf{p}_1-\frac{m_c}{m_c+m_u+m_d}\mathbf{p}_{\Lambda_c},\mathbf{p}_2-\frac{m_d}{m_c+m_u+m_d}\mathbf{p}_{\Lambda_c},
\mathbf{p}_3-\frac{m_u}{m_c+m_u+m_d}\mathbf{p}_{\Lambda_c})$ and
$\Psi^{tot}_{\Sigma}=\delta(\mathbf{p}_4+\mathbf{p}_5+\mathbf{p}_6-\mathbf{p}_{\Sigma})
\phi(\mathbf{p}_4-\frac{m_u}{2m_u+m_s}\mathbf{p}_{\Sigma},\mathbf{p}_5-\frac{m_u}{2m_u+m_s}\mathbf{p}_{\Sigma},\mathbf{p}_6-\frac{m_s}{2m_u+m_s}\mathbf{p}_{\Sigma})$.

The spatial wave functions of the ground-state baryon and meson are

$$\phi(\mathbf{p}_1,\mathbf{p}_2,\mathbf{p}_3)=(\frac{3R_B^2}{\pi})^{3/2}\exp[-\frac{R_B^2}{6}\sum_{i,j}(\mathbf{p}_i-\mathbf{p}_j)^2];$$

$$\phi(\mathbf{p}_a,\mathbf{p}_b)=(\frac{R_M^2}{\pi})^{3/2}\exp[-\frac{R_M^2}{8}(\mathbf{p}_1-\mathbf{p}_2)^2].$$

\end{document}